\documentclass[11pt]{article}
\usepackage{graphicx}
\usepackage[margin=1.25in]{geometry}
\usepackage[usenames,dvipsnames]{color}
\usepackage{url}
\usepackage[colorlinks = true,
            linkcolor = blue,
            urlcolor  = blue,
            citecolor = blue,
            anchorcolor = blue]{hyperref}
\usepackage{authblk}

\newenvironment{Abstract}{\begin{quotation} \begin{center}
                       ABSTRACT
     \end{center}\bigskip  }{\end{quotation}}

\newcommand\snowmass{\begin{center}\rule[-0.2in]{\hsize}{0.01in}\\\rule{\hsize}{0.01in}\\
\vskip 0.1in Submitted to the  Proceedings of the US Community Study\\ 
on the Future of Particle Physics (Snowmass 2021)\\ 
\rule{\hsize}{0.01in}\\\rule[+0.2in]{\hsize}{0.01in} \end{center}}

\begin{document}

%\pubblock

\title{Belle~II grid-based user analysis}

\bigskip 

% Authors -------------------------------------------------
  \author[1]{J.~V.~Bennett}\affil[1]{University of Mississippi, University, Mississippi 38677}
  \author[1]{J.~Guilliams}
  \author[2]{M.~Hern\'andez~Villanueva}\affil[2]{Deutsches Elektronen--Synchrotron, 22607 Hamburg, Germany}
  \author[3]{D.~E.~Jaffe}\affil[3]{Brookhaven National Laboratory, Upton, New York 11973}
  \author[3]{P.~J.~Laycock}
  \author[1]{A.~Panta}
  \author[3]{C.~Serfon}
  \author[4]{I.~Ueda}\affil[4]{High Energy Accelerator Research Organization (KEK), Tsukuba 305-0801, Japan}

\date{}

\normalsize
\maketitle

\vspace{-0.30in}

\noindent {\large \bf Corresponding Author:} \\
\noindent {J.~V.~Bennett (University of Mississippi), jvbennet@olemiss.edu} \\

 \begin{Abstract}
\noindent The Belle~II experiment at the SuperKEKB accelerator is a next-generation
B-factory aiming to collect 50 ab$^{-1}$, about 50 times the data collected
at Belle, to study rare processes and make precision measurements that may
expose physics beyond the Standard Model. Corresponding to roughly 100 PB of
storage for raw data, plus dozens of PBs per year for Monte Carlo (MC) and
analysis data, these massive samples require careful planning for the storage,
processing, and analysis of data. This white paper notes some of the challenges
that await grid-based user-analysis at the intensity frontier and invites
further discussion and exploration to improve the tools and techniques necessary 
to leverage the massive data samples that will be available at Belle~II as part of 
the Snowmass process.
\end{Abstract}

\snowmass

Physics analysis at the intensity frontier will face significant challenges over the coming decades, as massive data samples must be efficiently prepared, stored, and made accessible for analysts. Many of these challenges are common to multiple experiments. For example, collaborations face the prospect of working with increasingly larger data volumes using the same or fewer resources. Computing models must be reevaluated to make better use of available computing power and storage space. Several possible solutions to these challenges are also common, such as modifying software to run on GPU clusters and using streaming input. Other challenges and solutions are experiment-specific and will require innovation and careful planning. Here we note some of the challenges that will be faced by the Belle~II experiment and other intensity frontier experiments, along with some topics on which the broader community should engage.

The Belle~II experiment at the SuperKEKB asymmetric-energy e$^{+}e^{-}$ collider in Tsukuba, Japan, intends to collect 50 ab$^{-1}$ over the next decade. While data taking is on-going, Belle~II is a relatively young experiment. In this sense, it benefits from the ability to define data structures and design software while considering long-term implications and challenges. However, it also means that many of those challenges are far in the future and may not take a primary focus in the near term. Synergies with other experiments and reflection on workflows and tools that have been implemented elsewhere present opportunities to be better prepared for the challenges of the future.

\textbf{One of the major advantages of the Belle~II experiment with regard to long-term user analysis is the data format and declarative analysis language}~\cite{basf2}. The declarative approach means that users say “what” they want without specifying “how” they get it.  This allows software implementation to evolve with changes both in the underlying computing architectures (e.g. GPUs) and in the computing model (e.g. the elimination of intermediate data products can be made transparent to the user).

As in most HEP experiments, the first analysis step following event reconstruction is data reduction. Tracking and other information is used to build data structures that include all of the information characterizing the particles in each event. This information is used to skim data and MC samples and store only selected events of interest for a particular analysis. These skimmed files include the same data structures used as input as well as the particle-level data structures. In this way, analysts only have to run over events that pass the minimum criteria of the associated skim and need not recreate the particle-level information, greatly reducing the necessary computing power, though at the cost of about 30\% greater storage needs. Thanks to the declarative analysis language, individual analysts need only write high-level code, meaning that this computing model choice can be changed with minimal disruption to the analysts themselves, while the work done to accommodate these changes can be centralized and performed by experts.

While the data format has been constructed with considerations of large scale and limited resources in mind, it will be important to review the content and usage of analysis files and to determine a final output type suitable for long term data preservation, taking into account the evolving size and resource requirements. One possible way to keep the maximal amount of information would be to store the reconstructed output for all events that do not appear in at least one skimmed sample. In this way, the reconstructed information for every Belle~II event will be retained and analysts may pursue any analysis that is not covered by an analysis skim.

Belle~II estimates that storage of data from an integrated luminosity of 50 ab$^{-1}$ will require about 14 PB for skimmed data and about 16 PB for MC, assuming a sample equivalent to the acquired integrated luminosity and a single replica of each sample. This is at least an order of magnitude larger than the size of data preserved by previous experiments, like CDF, D0 and BaBar. For reference, the January 2021 capacity of the CERN Data Center was 558 PB and 222 PB for tape and disk, respectively. Multiple copies of the Belle~II reconstructed data should be preserved to avoid data loss from hardware issues. Preservation of two copies of the final output samples is currently estimated to require about 60 PB of storage.  \textbf{While this data volume is not challenging when compared to HL-LHC scales, it should be noted that it corresponds to 10$^{12}$ events which is a significant data management challenge. The per-event size of around 10kB means that this final Belle~II dataset is quite similar to the analysis datasets of the ATLAS and CMS experiments.}~\cite{hllc}

After the shutdown of the experiment, Belle~II collaborators will be required to seek resources for storage, computing, and networking from funding agencies to exploit grid-based analysis at all or a fraction of Belle~II grid sites. Since a grid-based system entails significant sustained effort to ensure security, it may be necessary to consider reducing the number of Belle~II sites or even to migrate to a non-grid system. A centralized (localized) computing model benefits from easier data management, lower maintenance, and easier user management efforts than a grid-based distributed computing model. However, a centralized model will be less scalable in some ways, though opportunistic use of resources allocated to other experiments may be possible, and certain sites may not be accessible to users from some regions. If national laboratory partners are providing central services for other experiments, it may be possible to continue running some services for Belle~II on a best-effort basis. Otherwise, it may be necessary to convert conditions services to read-only mode, which would prevent analysts from recording their own analysis conditions. \textbf{Broader discussions of the feasibility of long-term data analysis that relies on central services would be valuable to the HEP community. In particular, identifying services that would benefit from common use by many experiments would aid long-term data preservation and analysis plans.}

The Belle~II analysis software framework (basf2) is organized in modules written in C++, with a user interface implemented in Python 3~\cite{basf2}. The framework depends on external libraries and several software packages that include HEP-specific tools such as Geant4, ROOT, and physics generators, including EvtGen, Pythia, Tauola, etc. \textbf{While the analysis framework is currently supported and maintained by the collaboration, considerations must be made to ensure usability of the software by analysis after the shutdown of the experiment.} One solution may be providing a container image that includes the settings, system tools, and libraries needed to execute basf2. Such a solution also provides a minimal maintenance effort in the long term and ensures the compatibility of the image with the recent versions of containerization platforms.

It will be important to consider the long-term implications during development of grid-based user analysis tools when possible. One significant challenge is the scalability of tools for intensity frontier experiments as data volumes grow. Even relatively simple tasks like discovering datasets and submitting user jobs must have a scalable solution. At Belle~II, this is achieved by integrating grid-based user tools with Rucio, which resolves a large number of file paths embedded in a single container to reduce the time required for job submission. To simplify the task for users, “collections” of datasets are defined by the data production team. Users need only provide the collection name as an argument at submission time. Another challenge is the resolution of metadata at job submission and generally dealing with metadata for large data volumes. Resolving metadata on the server side improves the situation, but a more scalable approach may be required to avoid bottlenecks. Significant challenges are also present after job submission, for example a massive job submission with faulty code resulting in wasted resources. At Belle~II, a small set of “scout jobs” are required to succeed before full submission of jobs for a given project. Information from these scout jobs may also be used to identify job parameters such as CPU time and output file size. With additional development, these parameters can be used to improve the scalability of user analysis to cope with limited CPU and storage resources as data volumes increase. As data volumes become very large, scout jobs could even be used to provide users with a quick source of information on some set of distributions. Upon completion of user jobs, massive data samples will require scalable tools for asynchronous replication, deletion, and other operations. Integration of dedicated tools with Rucio is vital to these efforts and allows access to community-wide tools and workflows. Consideration should also be given to situations like file aggregation, i.e. production of many small files by users, that may not be easily addressed with Rucio.

In summary, the massive data samples that are currently being collected at intensity frontier experiments like Belle~II will require sophisticated tools and workflows to cope with limited resources. Future-minded, scalable development will enable physics analysis and data-mining of massive data samples well after the end of life of these experiments. Community-wide discussions will be vital to making this a reality.

%%%%%%%%%%%%%%%%%%%%%%%%%%%%%%%%%%%%%%%%%%

%  If you would like to use BibTEX for the bibliography, please feel free to do so.  It is not required.

%  To use BibTeX,

%    1.  uncomment the following two lines, 
%    2.  comment out everything below from  \begin{thebibliography}{99}   to \end{thebibliography).
%    3.  create the file  myreferences.bib, and process this file in the usual way

%\bibliographystyle{JHEP}
%\bibliography{myreferences}  % file myreferences.bib

%%%%%%%%%%%%%%%%%%%%%%%%%%%%%%%%%%%%%%%%%

\end{document}